# Growth window and possible mechanism of millimeter-thick single-walled carbon nanotube forests

Kei Hasegawa[1], Suguru Noda[1,*], Hisashi Sugime[1], Kazunori Kakehi[1], Shigeo Maruyama[2] and Yukio Yamaguchi[1]

[1] Department of Chemical System Engineering, School of Engineering, The University of Tokyo, 7-3-1 Hongo, Bunkyo-ku, Tokyo 113-8656, Japan
[2] Department of Mechanical Engineering, School of Engineering, The University of Tokyo, 7-3-1 Hongo, Bunkyo-ku, Tokyo 113-8656, Japan
Corresponding author. E-mail address: noda@chemsys.t.u-tokyo.ac.jp

Our group recently reproduced the water-assisted growth method, so-called "super growth", of millimeter-thick single-walled carbon nanotube (SWNT) forests by using $C_2H_4$/ $H_2$/ $H_2O$/ Ar reactant gas and Fe/ $Al_2O_3$ catalyst. In this current work, a parametric study was carried out on both reaction and catalyst conditions. Results revealed that a thin Fe catalyst layer (about 0.5 nm) yielded rapid growth of SWNTs only when supported on $Al_2O_3$, and that $Al_2O_3$ support enhanced the activity of Fe, Co, and Ni catalysts. The growth window for the rapid SWNT growth was narrow, however. Optimum amount of added $H_2O$ increased the SWNT growth rate but further addition of $H_2O$ degraded both the SWNT growth rate and quality. Addition of $H_2$ was also essential for rapid SWNT growth, but again, further addition decreased both the SWNT growth rate and quality. Because $Al_2O_3$ catalyzes hydrocarbon reforming, $Al_2O_3$ support possibly enhances the SWNT growth rate by supplying the carbon source to the catalyst nanoparticles. The origin of the narrow window for rapid SWNT growth will also be discussed.

## 1. INTRODUCTION

Single-walled carbon nanotubes (SWNTs) have unique mechanical and electrical properties, and many applications for them have been proposed and researched. To realize practical applications, mass production of SWNTs is essential, and various catalytic chemical vapor deposition (CCVD) methods have been developed to achieve this mass production. There are two types of CCVD; one involving nanoparticle catalysts suspended in the gas phase and the other involving nanoparticle catalysts supported on substrates. A gas-phase production process, the so-called "HiPco process", is the first process to be used in the mass production of SWNTs.[1] Recently, remarkable progress has been made in CCVD processes using supported catalysts. Submicrometer-thick films of randomly-oriented SWNTs have been the typical product when supported catalysts are used. In 2003, vertically aligned single-walled carbon nanotubes (VA-SWNTs) were realized[2] by using alcohol catalytic CCVD (ACCVD).[3] VA-SWNTs have now been achieved using several CVD methods and conditions.[4-7] As a result, CCVD from substrates now has potential as a process in the mass production of SWNTs.

Among those growth methods, the water-assisted method, so-called "super growth",[4] realized an outstanding growth rate of a few micrometers per second, leading to millimeter-thick VA-SWNTs forests. However, many research groups have failed in reproducing "super growth", and the underlying mechanism of the growth rate enhancement by water remained unclear. By using our combinatorial method for catalyst optimization,[8,9] we recently reproduced the "super growth" method and showed the important role of catalyst supports.[10] In this current work, by doing a parametric study, we report in detail the effect of the catalyst and reaction conditions determined, and discuss the novel mechanism essential for rapid growth VA-SWNTs.

## 2. EXPERIMENTAL

Catalysts were prepared on $SiO_2$ substrates by sputter-deposition. An $Al_2O_3$ layer was formed by depositing 15-nm-thick Al on a substrate and then exposing the layer to ambient air. Fe was deposited on Al$_2$O$_3$ layers or directly on SiO$_2$ substrates. For a separate experiment, gradient-thickness profiles of Fe were formed by using combinatorial masked deposition (CMD) method previously described.[9] The catalysts were set in a tubular CVD reactor (22 mm in diameter and 300 mm in length), heated to a target temperature (typically 1093 K), and kept at that temperature for 10 min while being exposed to 27 kPa H$_2$/75 kPa Ar at a flow rate of 500 sccm, to which H$_2$O vapor was added at the same partial pressure as for the CVD condition (i.e., 0 to 0.03 kPa). During this heat treatment, Fe forms nanoparticles of a certain diameter and areal density depending on the initial Fe thickness.[8] After the heat treatment, CVD was carried out by switching the gas to C$_2$H$_4$/ H$_2$/ H$_2$O / Ar. The standard condition was 8.0 kPa C$_2$H$_4$/ 27 kPa H$_2$/ 0.01 kPa H$_2$O/ 67 kPa Ar at 500 sccm at 1093 K for 10 min. The samples after CCVD were analyzed by using transmission electron microscopy (TEM) (JEOL JEM-2000EX), field emission scanning electron microscopy (FE-SEM) (Hitachi S-900), and micro-Raman scattering spectroscopy (Seki Technotron, STR-250) with an excitation wavelength at 488 nm.

## 3. RESULTS AND DISCUSSION
### 3.1. Standard condition of "super growth"

Figure 1a shows a photograph of CNT samples grown by a combinatorial catalyst library under the standard condition. Millimeter-thick vertically aligned CNTs (VA-CNTs) were grown at regions where the Fe thickness was 0.4 nm or more. The maximum thickness of VA-CNTs was 1.2 mm at a Fe thickness of 0.5 nm. The thickness of VA-CNTs decreased when Fe thickness exceeded 0.5 nm.

Figure 1b shows TEM images of CNTs grown under the same condition as Fig. 1a on substrates with uniform Fe thicknesses of 0.5 and 1.0 nm. SWNTs with a diameter around 4 nm mainly grew for 0.5-nm-thick Fe catalyst, whereas thicker CNTs grew for 1.0-nm-thick Fe catalyst. This difference in CNTs is because a thicker initial Fe layer yields larger Fe nanoparticles,[8] indicating a narrow VA-SWNTs growth window for the initial Fe thickness.

Figure 1c shows Raman spectra of the same CNT sample as Fig. 1a taken at Fe thicknesses of 0.5, 0.8, and 1.0 nm. Sharp, branched G-bands with a small D-band and peaks of radial breathing mode (RBM) were detected, indicating the existence of SWNTs. The G/D peak area ratios were smaller for thicker Fe layers ($\geq$ 1 nm), because multi-walled CNTs (MWNTs) became the main product when a thicker Fe layer was used as catalyst.

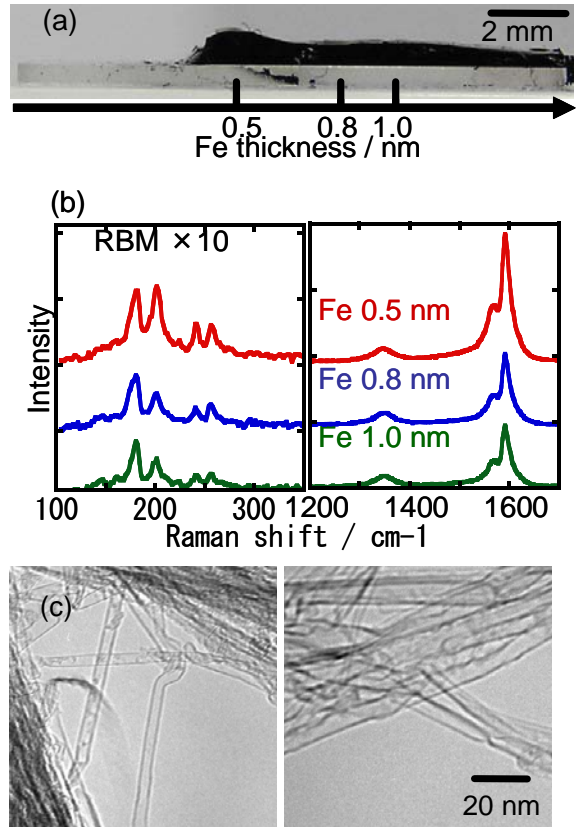

Fig. 1 CNTs grown on the Fe/Al$_2$O$_3$ catalyst library under the standard condition (8.0 kPa C$_2$H$_4$/ 27 kPa H$_2$/ 0.010 kPa H$_2$O/ 67 kPa Ar at 1093 K for 10 min). (a) Side view photograph of CNTs grown on the combinatorial catalyst library. (b) Raman spectra of the same sample at Fe thickness of 0.5, 0.8 and 1.0 nm. Intensity at the low wavenumber region (< 300 cm$^{-1}$) was 10x magnified. (c) TEM images of CNTs grown on substrates with uniform Fe thickness of 0.5 nm (left) and 1.0 (right) nm.

These figures show that "super growth" was achieved in this work. The growth temperature of the standard condition of this work is higher than that of the original "super growth"[4] because both the CNT thickness and the G/D ratio increased with increasing growth temperature.

### 3.2. Effects of catalyst metals and supports

Effects of catalyst metals and their supports were investigated next. Figure 2a shows top-view photographs of CNT samples grown on the Fe/Al$_2$O$_3$, Co/Al$_2$O$_3$, and Ni/Al$_2$O$_3$ combinatorial catalyst libraries under the standard condition. The surfaces of both Fe/Al$_2$O$_3$ and Co/Al$_2$O$_3$ libraries became black at regions where the Fe thickness was 0.4 nm or more. On the other hand, the surface of Ni/Al$_2$O$_3$ was somewhat darkened only at the relatively thin Fe region around 0.5 nm. Figure 2b shows cross-sectional SEM images of these samples. About 200-μm-thick VA-CNTs grew for 0.5-nm-thick Co and about 0.4-μm-thick VA-CNTs grew for 0.5-nm-thick Ni..

Figure 2c shows top-view photographs of combinatorial catalyst libraries of Fe/SiO$_2$, Co/SiO$_2$, and Ni/SiO$_2$ after CVD under the standard condition. The results were completely different from those libraries with Al$_2$O$_3$ support layer (i.e., Fe/Al$_2$O$_3$, Co/Al$_2$O$_3$, and Ni/Al$_2$O$_3$); VA-CNTs did not form on any library and only part of the surface of Fe/SiO$_2$ for the thickness range from 0.4 to 0.6 nm darkened slightly compared with the library before CVD. As for Co/SiO$_2$ and Ni/SiO$_2$, negligible change appeared in color. Figure 2d shows the Raman spectra for the Fe/ SiO$_2$ sample in Fig. 2b at Fe thickness of 0.5 nm, indicating growth of SWNTs. Fe catalyst supported on SiO$_2$ can actually grow CNTs, including SWNTs, although the CNT yield is much smaller (submicrometer thickness or less) than that grown using catalysts supported on Al$_2$O$_3$ (up to millimeter thickness), regardless of the thickness of the catalyst.

Next, we examined if an Al$_2$O$_3$ catalyst support layer actually acts as Al$_2$O$_3$ rather than metallic Al. Another catalyst library was prepared by first depositing 15-nm-thick Al on SiO$_2$, then exposing the layer to ambient air, then oxidizing the layer under air at 973 K for 5 min, and finally depositing a gradient thickness profile of Fe on the layer at ambient temperature. Figure 3a shows photographs of the Fe/ Al$_2$O$_3$ catalyst libraries before CVD. In the thick Fe region (≥ 2 nm), both libraries were metallic silver in color. In the thinner region (< 2 nm), the Al layer only exposed to air was gray metallic in color, indicating the existence of metallic Al, whereas the Al layer oxidized at 973 K was completely transparent, indicating the complete oxidation of the Al layer. Then, CVD was carried out on these libraries. Figure 3b shows the photograph after CVD of the library with the Al layer oxidized at high temperature. The growth of the VA-CNT films was similar to that without high-temperature oxidation of the Al layer (Fig. 1a), indicating that the surface of the Al layer was oxidized by just being exposed to air, and that the layer acted as an Al$_2$O$_3$ support.

Fe nanoparticles can grow SWNTs on either SiO$_2$ or Al$_2$O$_3$ supports, but they need to be supported on $Al_2O_3$ to achieve rapid growth of VA-CNTs at a rate of micrometers per second. In addition, $Al_2O_3$ support enhances the CNT growth by other catalysts (i.e., Co and Ni). $Al_2O_3$ support layer should have an essential role in growing CNTs.

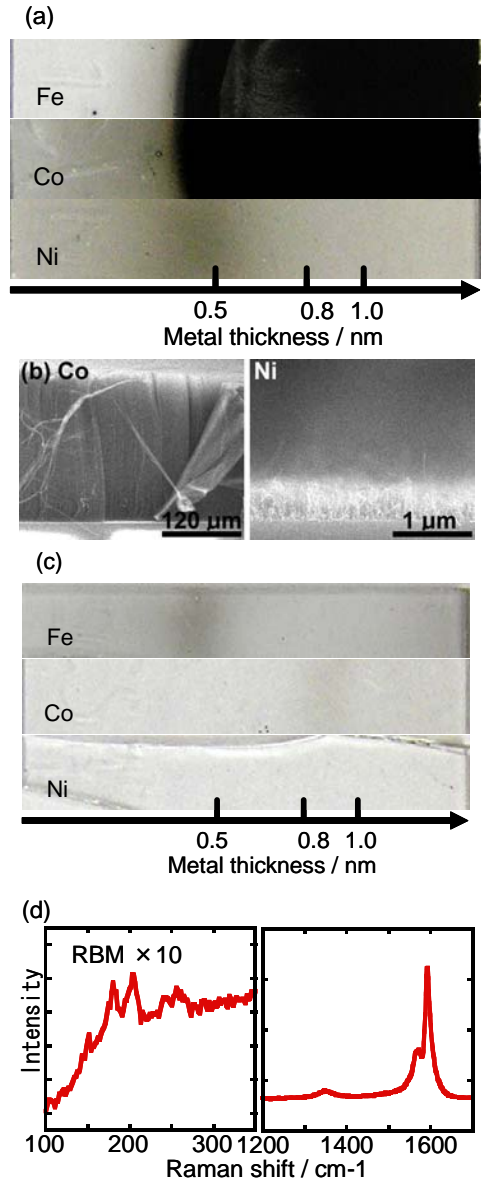

Fig. 2 Effects of catalyst metals and supports on CNT growth under the standard condition. (a,c) Top-view photographs of CNTs samples grown by combinatorial catalyst libraries prepared on Al$_2$O$_3$ (a) and SiO$_2$ (c). All of the catalyst metals (i.e. Fe, Co, Ni) had the same thickness profiles between 0.2 nm (left) and 3 nm (right). (b) Cross-sectional FE-SEM images of the CNTs samples grown by 0.5-nm-thick Co and Ni catalysts prepared on Al$_2$O$_3$. (d) Raman spectrum of the CNTs samples grown by 0.5-nm-thick Fe catalyst prepared on SiO$_2$ supports. Intensity at the low wavenumber region (<300 cm$^{-1}$) was 10x magnified.

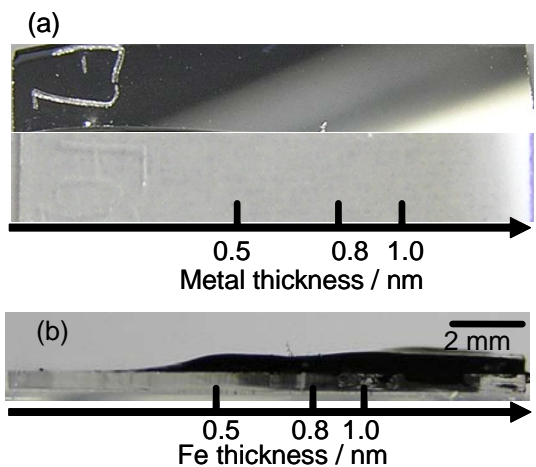

Fig. 3 Effect of the preparation process of $Al_2O_3$ layer. (a) Photographs of $Fe/Al_2O_3$ catalyst libraries before CVD. $Al_2O_3$ layer was formed by oxidizing 15-nm-thick Al layer only by exposure to ambient air (top) and then by oxidizing in air at 973 K (bottom). (b) Photograph of CNT grown on the $Fe/Al_2O_3$ library with $Al_2O_3$ layer oxidized at 973 K (same library as Fig. 3a, bottom).

3.3. Effects of $H_2O$ and $H_2$

Figure 4a shows the thickness profiles of CNTs grown on $Fe/Al_2O_3$ catalyst libraries under the standard condition, except that the amount of added $H_2O$ (i.e. partial pressure) was varied. Without $H_2O$, CNTs grew only at the relatively thin Fe region (0.3-1.0 nm) and the maximum thickness of VA-CNTs was 0.7 mm at the 0.5 nm-thick Fe region. Addition of 0.010 kPa $H_2O$ enhanced the growth, especially at the thick Fe region (> 0.7 nm), and the maximum VA-CNT thickness increased to 1.0 mm at the 0.5-nm-thick Fe region. Further $H_2O$ addition (0.030 kPa) inhibited CNT growth at the thin Fe region (0.3- 0.6 nm) where SWNTs grew at the lower $H_2O$ partial pressures (≤ 0.010 kPa). Figure 4b shows the G/D ratios for each CVD condition at Fe thickness of 0.5, 0.8, and 1.0 nm. Slight addition of $H_2O$ (0.010 kPa) did not affect the G/D ratio at the thin Fe region (0.5 nm), but decreased the G/D ratio at the thicker regions (0.8 and 1.0 nm). Further addition of $H_2O$ (0.030 kPa) significantly decreased the G/D ratio at all regions. These results show that proper amount of $H_2O$ enhances the growth rate of CNTs, but excessive $H_2O$ decreases the growth rate of CNTs from small nanoparticles (i.e. small Fe thickness) and degrades the quality of CNTs possibly by oxidation. In conclusion, excess $H_2O$ totally inhibits "super growth" of SWNTs.

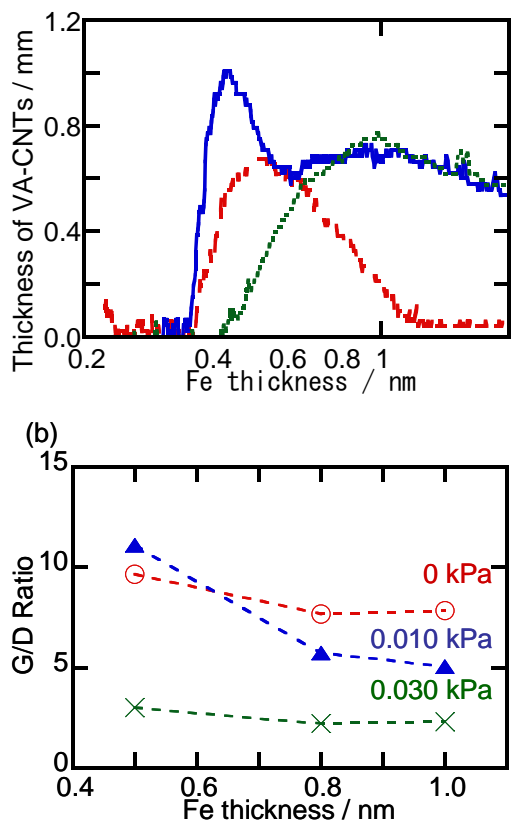

Fig. 4 Effect of $H_2O$ on CNT growth. CVD was carried out by using $Fe/Al_2O_3$ combinatorial catalyst libraries under the standard condition except the amount of added $H_2O$ (partial pressure) was varied. (a) Relationship between thickness of VA-CNTs and Fe thickness at different $H_2O$ pressures. (b) Relationship between G/D ratio of Raman spectra and Fe thickness at different $H_2O$ pressures.

Figure 5a shows thickness profiles of CNTs grown under the standard condition except that the amount of added $H_2$ (i.e. partial pressure) was varied. When a lower amount of $H_2$ was added (2.7 kPa), VA-CNTs became much thinner at Fe thickness of 0.7 nm or less. When a large amount of $H_2$ was added (54 kPa), CNTs grew at any Fe thickness at a reduced VA-CNT thickness (around 0.4 mm). Figure 5b shows the G/D ratios for each CVD condition at Fe thickness of 0.5, 0.8, and 1.0 nm. The G/D ratios decreased in either case of lower (2.7 kPa) and higher (54 kPa) $H_2$ was added. These results indicate that an optimum amount of $H_2$ is needed for rapid growth of VA-SWNTs of relatively good quality.

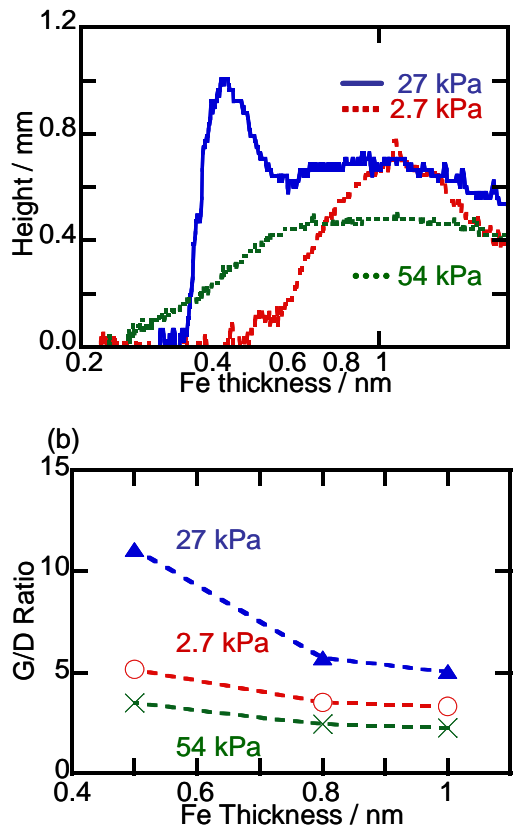

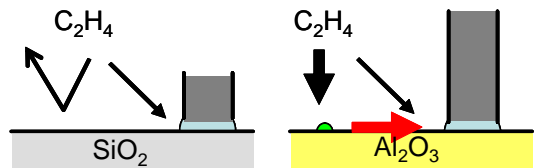

Fig. 6 Schematic of enhancement mechanism of SWNT catalytic growth by $Al_2O_3$ support.

Fig. 5 Effect of $H_2$ on CNT growth. CVD was carried out by using $Fe/Al_2O_3$ combinatorial catalyst libraries under the standard condition except the amount of added $H_2$ (partial pressure) was varied. (a) Relationship between thickness of VA-CNTs and Fe thickness at different $H_2$ pressures. (b) Relationship between G/D ratio of Raman spectra and Fe thickness at different $H_2$ pressures.

3.4. Possible mechanism of rapid growth of VA-SWNTs

Based on the results discussed above, we propose three necessary conditions for "super growth" of SWNTs. The first condition is that the Fe catalyst needs to be thin enough, about 0.5 nm, so that the catalyst nanoparticles are small enough to grow SWNTs with small diameters. The second condition is that the catalyst nanoparticles need to be supported on $Al_2O_3$. The third is that the partial pressures of both $H_2O$ and $H_2$ need to be carefully adjusted; these gases are essential but excessive amounts degrade the growth rate and/or the quality of SWNTs.

$Al_2O_3$ and its related materials have long been used as catalysts for decomposition and dehydrogenation of hydrocarbons.[11,12] Figure 6 shows a schematic of our proposed mechanism explaining the enhancement effect of $Al_2O_3$ on CNT growth. In this mechanism, $C_2H_4$ or its derivatives adsorb on $Al_2O_3$, diffuse over $Al_2O_3$ surface to catalyst nanoparticles, and then are incorporated in Fe nanoparticles. In contrast, when Fe is deposited directly on $SiO_2$, $C_2H_4$ can be decomposed only on Fe nanoparticles whose surface is largely covered with growing CNTs. As a result, the growth rate of CNTs on $Al_2O_3$ is much faster than that on $SiO_2$. In addition, the catalytic activity of $Al_2O_3$ strongly depends on its crystalline structure.[11] $\alpha$-$Al_2O_3$, the most stable phase, has low activity,[11] and thus, rapid CNT growth does not occur when catalysts are supported on sapphire (i.e. monocrystalline $\alpha$-$Al_2O_3$).

Next, we discuss why the window of "super growth" for SWNTs is narrow. One reason might be due to the catalyst deactivation mechanism, i.e. coking of either $Al_2O_3$ surface or Fe nanoparticles. In the dehydrogenation process of hydrocarbons,[11] $Al_2O_3$ easily loses its catalytic activity due to carbon deposition, and thus, $H_2O$ vapor has long been used to remove carbon byproducts. $H_2$ also keeps the $Al_2O_3$ surface reactive by controlling the balance between dehydrogenation and hydrogenation of carbon surface species, as is known in the hydrocracking process.[13] In contrast, excessive $H_2$ inhibits the growth of SWNTs by hydrogenating $C_2H_4$-derivatives adsorbed on $Al_2O_3$. Concerning Fe nanoparticles, when the incoming flux of carbon into Fe nanoparticles increases, carbon in Fe nanoparticles will be highly supersaturated, resulting not only in increased SWNT growth rate but also in graphite formation on the surface of the nanoparticles. The fewer walls and larger free energy of SWNTs than MWNTs may make the degree of supersaturation larger for nanoparticles growing SWNTs than those growing MWNTs. This may be the reason why the "super growth" window for SWNTs is such narrow compared with MWNTs. (Note that MWNTs grow rapidly under a wide window of reaction conditions if catalyst nanoparticles are supported on $Al_2O_3$). In conclusive, two conditions are needed to sustain the rapid growth of SWNTs: first, the partial pressures of $C_2H_4$, $H_2$, and $H_2O$ need to be balanced to suppress coking of $Al_2O_3$, and second, the incoming flux of carbon into Fe nanoparticles must not be too large, that means nanotubes must not grow too rapid, to prevent carbonization of Fe nanoparticles.

**4. CONCLUSION**

Rapid growth of VA-SWNTs, namely, a few micrometers per second, or so-called "super growth", was reproduced in this study, and the growth window was clarified. The standard condition of this work was 8.0 kPa $C_2H_4$/ 27 kPa $H_2$/ 0.01 kPa $H_2O$/ 67 kPa Ar at 500 sccm at 1093 K for 10 min using a tubular CVD reactor (22 mm in diameter and 300 mm in length). Results showed that for the rapid growth of SWNTs, small nanoparticles formed from a thin Fe layer (about 0.5 nm) need to be supported on $Al_2O_3$ and that only the optimal amounts of $H_2O$ and $H_2$ should be used. We proposed a novel mechanism by which $Al_2O_3$ enhances the growth rate of SWNTs, and offered a simple explanation of the effect of $H_2O$ and $H_2$ on the growth rate. Namely, $Al_2O_3$ supports supply a carbon source to Fe nanoparticles, and $H_2O$ and $H_2$ prevent catalyst deactivation by keeping the $Al_2O_3$ surface reactive and by balancing the carbon fluxes among the gas-phase, $Al_2O_3$ support, Fe nanoparticles, and growing SWNTs.

**Acknowledgements:**
The authors thank Z. Zhang for her help in the Raman measurements. This work is financially supported in part by the Grant-in-Aid for Young Scientists (A), 18686062, 2006, from the Ministry of Education, Culture, Sports, Science and Technology (MEXT), Japan.